\documentclass[a4paper, amsfonts, amssymb, amsmath, 
reprint, showkeys, dvipsnames, nofootinbib, twoside,superscriptaddress,aps,prd,longbibliography]{revtex4-1}

\usepackage[english]{babel}
\usepackage[utf8]{inputenc}
\usepackage[colorinlistoftodos, color=green!40, prependcaption]{todonotes}

\usepackage{amsthm}
\usepackage{mathtools}
\usepackage{graphicx}
\usepackage[left=23mm,right=13mm,top=35mm,columnsep=15pt]{geometry} 
\usepackage{adjustbox}
\usepackage{placeins}
\usepackage[T1]{fontenc}
\usepackage{lipsum}
\usepackage{csquotes}
\usepackage[colorlinks = true, linkcolor = blue,
            urlcolor  = blue,citecolor = blue,
            anchorcolor = blue]{hyperref}% add hypertext capabilities
\usepackage[noabbrev]{cleveref}
\usepackage{siunitx}
\usepackage[normalem]{ulem}
\begin{document}
\title{Automated control and optimisation of laser driven ion acceleration}

\newcommand{\JAI}{The John Adams Institute for Accelerator Science, Imperial College London, London, SW7 2AZ, UK}

\newcommand{\GOLP}{GoLP/Instituto de Plasmas e Fus\~{a}o Nuclear, Instituto Superior T\'{e}cnico, U.L., Lisboa 1049-001, Portugal}

\newcommand{\CLF}{Central Laser Facility, STFC Rutherford Appleton Laboratory, Didcot OX11 0QX, UK}

\newcommand{\UCL}{Department of Physics and Astronomy, University College London, London WC1E 6BT, UK}

\newcommand{\LMU}{Fakult\"at f\"ur Physik, Ludwig-Maximilians-Universit\"at M\"unchen, D-85748 Garching, Germany}
\newcommand{\MPQ}{Max-Planck-Institut f\"ur Quantenoptik, Hans-Kopfermann-Str. 1, D-85748 Garching, Germany}

\newcommand{\DESY}{Deutsches Elektronen-Synchrotron DESY, Notkestr. 85, 22607 Hamburg, Germany}

\newcommand{\CI}{The Cockcroft Institute, Keckwick Lane, Daresbury, WA4 4AD, UK}

\newcommand{\LANCS}{Physics Department, Lancaster University, Lancaster LA1 4YB, UK}

\newcommand{\UMICH}{G\'erard Mourou Center for Ultrafast Optical Science, University of Michigan, Ann Arbor, MI 48109-2099, USA}

\newcommand{\ERGODIC}{Ergodic LLC, San Francisco, CA 94117, USA}

\newcommand{\SUPA}{Department of Physics, SUPA, University of Strathclyde, Glasgow G4 0NG, UK}

\newcommand{\LLNL}{Lawrence Livermore National Laboratory (LLNL), P.O. Box 808, Livermore, California 94550, USA}

\newcommand{\DLS}{Diamond Light Source, Harwell Science and Innovation Campus, Fermi Avenue, Didcot OX11 0DE, UK}

\newcommand{\YORK}{York Plasma Institute, School of Physics, Engineering and Technology, University of York, York YO10 5DD, UK}

\newcommand{\LUND}{Department of Physics, Lund University, P.O. Box 118, S-22100, Lund, Sweden}

\newcommand{\CHALMERS}{Department of Physics, Chalmers University of Technology, SE-41296 Gothenburg, Sweden}

\newcommand{\QUB}{School of Mathematics and Physics, Queen's University Belfast, BT7 1NN, Belfast UK}

\newcommand{\ELI}{ELI Beamlines Centre, Institute of Physics, Czech Academy of Sciences, Za Radnic\'{i} 835, 252 41 Doln\'{i} B\u{r}e\u{z}any, Czech Republic}

\newcommand{\SLAC}{SLAC National Accelerator Laboratory, 2575 Sand Hill Road, Menlo Park, California, USA}

\newcommand{\TUD}{Institut für Kernphysik, Technische Universität Darmstadt, Karolinenplatz 5, 64289 Darmstadt, Germany}

\newcommand{\SU}{Department of Applied Physics, Stanford University, Stanford, California 94305, USA}

\newcommand{\MSLAC}{Department of Mechanical Engineering, Stanford University, Stanford, California 94305, USA}
\newcommand{\UOA}{Department of Electrical and Computer Engineering, University of Alberta, Edmonton, AB, T6G1H9, Canada}

\newcommand{\Prague}{Faculty of Nuclear Sciences and Physical Engineering, Czech Technical University in Prague, Prague,
Czech Republic}

\author{B.~Loughran}
\affiliation{\QUB}

\author{M.~J.~V. Streeter}
\affiliation{\QUB}

\author{H.~Ahmed}
\affiliation{\CLF}

\author{S.~Astbury}
\affiliation{\CLF}

\author{M.~Balcazar}
\affiliation{\UMICH}

\author{M.~Borghesi}
\affiliation{\QUB}

\author{N.~Bourgeois}
\affiliation{\CLF}

\author{C.~B.~Curry}
\affiliation{\SLAC}
\affiliation{\UOA}

\author{S.~J.~D. Dann}
\affiliation{\CLF}

\author{S.~DiIorio}
\affiliation{\UMICH}

\author{N.~P. Dover}
\affiliation{\JAI}

\author{T.~Dzelzanis}
\affiliation{\CLF}

\author{O.~C.~Ettlinger}
\affiliation{\JAI}

\author{M.~Gauthier}
\affiliation{\SLAC}

\author{L.~Giuffrida}
\affiliation{\ELI}

\author{G.~D.~Glenn}
\affiliation{\SLAC}
\affiliation{\SU}

\author{S.~H.~Glenzer}
\affiliation{\SLAC}

\author{J.~S.~Green}
\affiliation{\CLF} 

\author{R.~J.~Gray}
\affiliation{\SUPA}

\author{G.~S.~Hicks}
\affiliation{\JAI}

\author{C.~Hyland}
\affiliation{\QUB}

\author{V.~Istokskaia}
\affiliation{\ELI}
\affiliation{\Prague}

\author{M.~King}
\affiliation{\SUPA}

\author{D.~Margarone}
\affiliation{\QUB}
\affiliation{\ELI}

\author{O.~McCusker}
\affiliation{\QUB}

\author{P.~McKenna}
\affiliation{\SUPA}

\author{Z.~Najmudin}
\affiliation{\JAI}

\author{C.~Parisua{\~{n}}a}
\affiliation{\SLAC}
\affiliation{\MSLAC}

\author{P.~Parsons}
\affiliation{\QUB}

\author{C.~Spindloe}
\affiliation{\CLF}

\author{D.~R. Symes}
\affiliation{\CLF}

\author{A.~G.~R. Thomas}
\affiliation{\UMICH}

\author{F.~Treffert}
\affiliation{\SLAC}
\affiliation{\TUD}

\author{N.~Xu}
\affiliation{\JAI}

\author{C.~A.~J. Palmer}
\affiliation{\QUB}
\email[Correspondence email address: ]{c.palmer@qub.ac.uk}

% \affiliation{School of Maths and Physics, Queen's University Belfast, University Road, Belfast, N. Ireland, U. K.}
% \affiliation{Central Laser Facility, STFC Rutherford Appleton Laboratory, Harwell Campus, Didcot, U. K.}
% \affiliation{Center for Ultrafast Optical Science, University of Michigan Engineering, 1221 Beal Ave., Ann Arbor, Michigan, U. S. A}
% \affiliation{Department of Physics, University of Strathclyde, 16 Richmond Street, Glasgow, U. K.}
% \affiliation{SLAC National Accelerator Laboratory, 2575 Sand Hill Road, Menlo Park, California, U. S. A.}
% \affiliation{ELI Beamlines, Za Radnic\'{i} 835, 252 41 Doln\'{i}  B\u{r}e\u{z}any, Czech Republic}
% \affiliation{Blackett Laboratory, Imperial College London, South Kensington Campus, London, U. K.}
\date{\today} 

\begin{abstract}

The interaction of relativistically intense lasers with opaque targets represents a highly non-linear, multi-dimensional parameter space. 
This limits the utility of sequential 1D scanning of experimental parameters for the optimisation of secondary radiation, although to-date this has been the accepted methodology due to low data acquisition rates.
High repetition-rate (HRR) lasers augmented by machine learning present a valuable opportunity for efficient source optimisation. 
Here, an automated, HRR-compatible system produced high fidelity parameter scans, revealing the influence of laser intensity on target pre-heating and proton generation.
A closed-loop Bayesian optimisation of maximum proton energy, through control of the laser wavefront and target position, produced proton beams with equivalent maximum energy to manually-optimized laser pulses but using only 60\% of the laser energy.
This demonstration of automated optimisation of laser-driven proton beams is a crucial step towards deeper physical insight and the construction of future radiation sources.

\end{abstract}


\maketitle

\section{Introduction} \label{sec:intro}

Laser-target interactions have been demonstrated to provide a highly versatile source of secondary radiation, of interest for many applications \cite{Brenner2015Laser-drivenAccelerators,Courtois2013CharacterisationRadiography} as well as the study of fundamental science \citep{Jaroszynski2006RadiationInteractions,Galy2009High-intensitySources,Zylstra2016}.
Specifically, laser-driven ion accelerators \cite{Daido2012ReviewApplications,Macchi2013RMP} have desirable characteristics pertaining to applications in medicine \cite{Bulanov2002OncologicalAccelerators}, material science \cite{Passoni2019AdvancedScience}, nuclear fusion \cite{Roth2001} and imaging \cite{Borghesi2002ElectricTechnique,Borghesi2001ProtonStudies,Mackinnon2004Protoninvited}. 
The need for stable, reproducible beams that can be tuned presents a necessary, yet challenging, goal towards the realisation of many of these applications \cite{Daido2012ReviewApplications}.

The most extensively researched mechanism driving laser-driven proton acceleration is sheath acceleration (often termed Target Normal Sheath Acceleration, or TNSA)\cite{Snavely2000}.  
Here, a laser pulse is focused to a relativistic intensity on the surface of a solid foil, ionising the material and heating electrons to MeV temperatures. 
As the accelerated electrons escape the rear target surface, they build an electrostatic sheath field which can reach $\gtrsim 1\,\si{TV/m}$ and accelerate ions to $\gg 1\,\si{MeV}$ energies over just a few microns \cite{RothIon}.  
Although target materials vary, the accelerated ions are typically dominated by protons from hydrocarbon surface contaminants which are preferentially accelerated due to their high charge-mass ratio \cite{Allen2004DirectFoils}.

Due to their dependence on the rear surface electrostatic sheath field, the specific characteristics of these MeV proton beams are strongly influenced by the laser-electron energy coupling, electron transport through the bulk of the target, and disruption to the target rear surface. 
Experiments have demonstrated the dependence of the electron and proton beam on various experimental control parameters (e.g. laser intensity and contrast or target thickness). 
Of particular importance is the plasma scale length at the front surface which affects the laser-electron coupling mechanisms \cite{Chopineau2019IdentificationPlasmas,MckennaEffectsTargets,Gray2014LaserGradients}.
In preplasmas with long scale lengths ($>100\,\si{\micro\meter}$) the laser beam has been observed to filament, subsequently reducing the coupling efficiency, while for optimal scale lengths the beam can undergo relativistic self-focussing effects that enhance laser energy coupling \citep{MckennaEffectsTargets,Gray2014LaserGradients,Esirkepov2014PrepulseTargets}.
The amplified stimulated emission (ASE) pedestal and prepulses, common to short pulse lasers, can pre-heat the target and lead to significant plasma expansion before the main pulse arrives.  For a given ASE pedestal duration, an optimal thickness exists at which this enhanced coupling and electron recirculation \cite{Mackinnon2002EnhancementPulses} will be advantageous for the acceleration process, while for thinner targets the inward travelling shock-wave launched by the rapid surface pre-heating can disrupt the accelerating sheath field at the rear surface \cite{Kaluza2004InfluenceExperiments, Lindau2005,Lundh2007b}. 
Radiative heating from x-rays, generated in the focus of a prepulse incident on the target front-side, can similarly induce rear-surface expansion of thin targets, impacting TNSA for interaction parameters in which the ASE induced shock may not reach the rear surface during the acceleration window \citep{Kaluza2004InfluenceExperiments}.

While broad trends within TNSA proton beams have been established, comparison of experimental results from different experiments highlights variation in measured beam parameters and is indicative of the nuanced relationship between laser parameters and proton beam characteristics. 
For example, while maximum proton energy has been observed to increase with laser intensity, the scaling follows a ${\sim}I^{1/2}$ dependence for long ($>300~\,\si{fs}$) duration laser pulses and a linear dependence ${\sim}I$ for ultra-short ($40-150\,\si{fs}$) pulses  \cite{Fuchs2005,Borghesi2008,Macchi2013RMP}. 
A number of numerical and experimental studies have explored the impact of laser pulse duration on maximum proton energy due to TNSA in interactions with moderate laser contrast.  These indicate an optimal pulse duration for proton acceleration associated with a fixed laser intensity or laser energy \cite{Oishi2005,Schreiber2006, Mora2003, Carrie2009} (e.g. optimal duration between $100-300\,\si{fs}$ for laser pulse energies of \SI{1}{J}).  The dependence is attributed to differences in laser energy to electron coupling efficiency and acceleration time relative to the rear surface expansion timescale.  For higher laser contrast, a similar trend is observed for targets with thicknesses of tens of microns \cite{Flacco}. 
More advanced temporal pulse shaping (e.g. shaping of the rising and falling edge of the pulse) has been explored recently for the interaction of high-contrast lasers with ultra-thin targets which indicate significant enhancement of proton maximum energies over those observed for best laser compression \cite{Tayyab2018EffectTargets,Ziegler2021ProtonSystem}.

With the proliferation of multi-Hz high power laser pulses \cite{Danson2019HPLSE}, and the development of HRR-compatible solid-density targetry \cite{Xu2022TapeTarget,Treffert2022Ambient-temperatureScience,Kim2016DevelopmentExperiments, Kraft2018FirstTarget, Schumacher2017LiquidAcceleration, Fraga2012CompactGeneration, Lemos2009DesignInteractions}, it is now possible to quickly obtain large datasets from laser-driven ion acceleration experiments.
This opens the possibility to perform extensive multi-dimensional parameter scans to elucidate the interdependence of different experimental control parameters, as well as to apply machine learning techniques to optimise ion beam properties - within complex multi-dimensional parameter spaces - in automated experiments \cite{Noaman-Ul-Haq2017,Shalloo2020, Jalas2021BayesianAccelerator} and simulations \cite{Dolier2022Multi-parameterSimulations, Djordjevic2021}.

Here, we describe the first experimental demonstration of real-time Bayesian optimisation of a laser-driven ion source, using a closed-loop algorithm. 
The fully automated control system operated the laser,  analysed the diagnostic results and made changes to the experiment control parameters. 
This enabled rapid and efficient optimisation of the accelerator performance through simultaneous tuning of up to six different input parameters, producing proton beams with equivalent peak energy using \SI{57}{\%} of the laser energy of the manually-optimised interaction.

\section{Experimental Setup} \label{sec:expt}

\begin{figure}[h!]
    \includegraphics[width=0.45\textwidth]{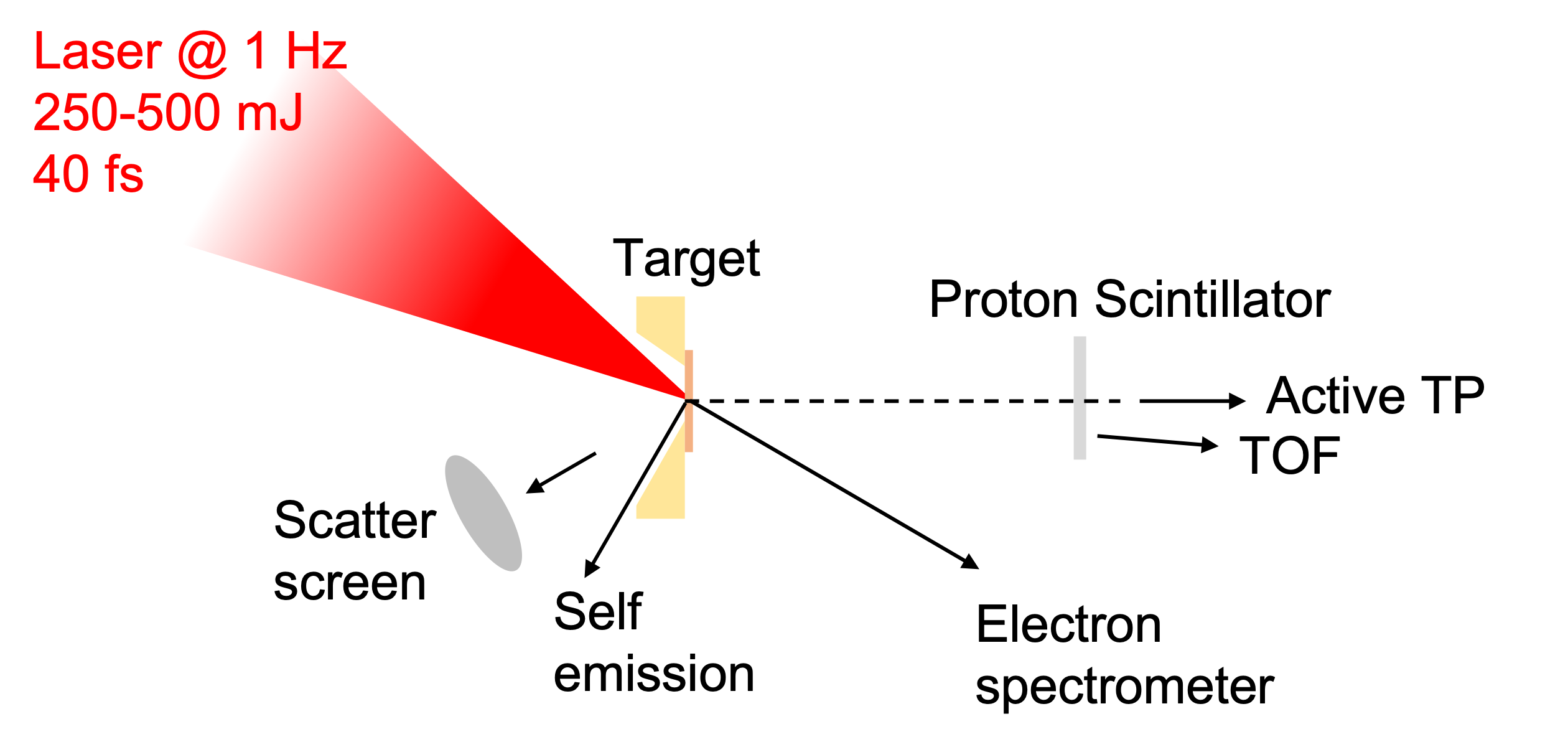}
    \caption{Illustration of the experimental setup, showing the orientation of the laser-plasma interaction and the main diagnostics. The laser was focused, with an f/2.5 90$^{\circ}$ diamond-turned OAP, to \SI{1.6}{\micro\meter} radius focal spot containing a median of $35\pm 3\%$ of pulse energy. The plane of the laser-plasma interaction was monitored by imaging self-emission at \SI{800}{nm} at $60^{\circ}$ to the laser propagation axis. %(see Appendix for details).  
    }
    \label{fig_setup}
\end{figure}

The experiment (see \cref{fig_setup} for setup) was performed at the Gemini TA2 facility, using a Ti:Sa laser which forms part of the Central Laser Facility at the Rutherford Appleton Laboratory. 
The laser pulses contained up to \SI{500}{mJ} in a transform-limited pulse duration FWHM of $\sim40\,\si{fs}$, with a central wavelength of \SI{800}{nm} and a FWHM bandwidth of \SI{30}{nm}.
The laser was focused to a high intensity ($I_L > 10^{19}\,\si{Wcm^{2}}$) using an $f/2.5$ off-axis parabolic mirror and interacted with the target at an angle of incidence of 30$^{\circ}$ with p-polarisation.
The target was Kapton tape of \SI{12}{\micro\meter}, spooled continuously during shots using a motorised tape drive \cite{Xu2022TapeTarget}. 

The interaction was diagnosed using a suite of particle diagnostics including a scintillator (EJ-440), positioned along the rear surface target normal, to measure the proton spatial profile, two point measurements of the proton energy spectrum using a time-of-flight (TOF) diamond detector \cite{Margarone2011FullDetectors} and fibre-coupled Thomson parabola spectrometer (at 3$^{\circ}$ to the target normal and along the target normal axis respectively), and a 0.15~T permanent magnet electron spectrometer in the laser-forward direction. 
The near-field of the specularly reflected laser light was also measured at the first and second harmonic of the drive laser.

The laser spectral phase was controlled by an acousto-optic programmable dispersive filter (DAZZLER) and measured using a small central sample of the compressed pulse and a SPIDER diagnostic. 
 The laser parameters (6 wavefront aberrations generated through Zernike polynomials, 2nd-, 3rd- and 4th-order temporal phase, energy and polarisation) and target position relative to the laser focus were controlled using a fully automated control and acquisition system.   This enabled data scans consisting of bursts of shots (up to 20) at fixed input values in parameter space.  Following a burst of shots, the control code performed analysis of  the measured data from the online diagnostics and adjusted laser or target parameters for the next burst. Although the controls enabled adjustment of the laser temporal pulse shape, the data presented here corresponds to pulse shapes close to best compression ($\sim 40\,\si{fs}$) with variations due to the day-to-day variation in laser tuning.

For high-energy, multi-Hz laser facilities, prolonged high-repetition rate operation can affect the laser pulse parameters leading to degradation in peak intensity \cite{Li2017DegradationGratings}.  
To ensure that our setup was not subject to these effects, measurements of laser parameters were made over periods of extended \SI{1}{Hz} operation. 
These measurements concluded that the effect of prolonged HRR operation on the quality of the temporal pulse shape was negligible, with the standard deviation of the random fluctuation in pulse FWHM measured as $\sim~3\,\si{fs}$ in over 1400 shots.

\section{Automated grid scans} \label{sec:results}

\begin{figure}[h!]
    \includegraphics[width=8.4cm]{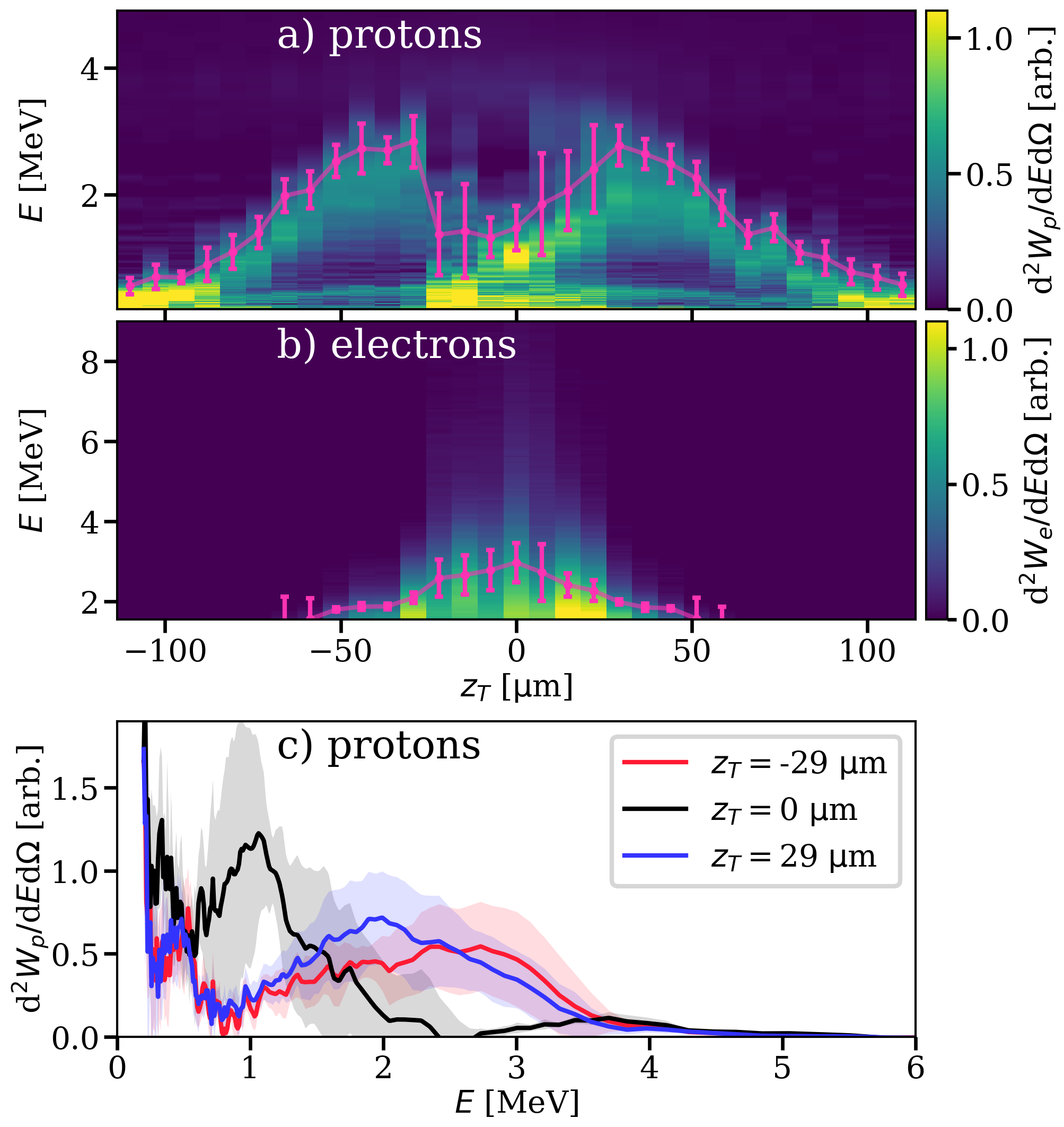}
    %20210521 run06
    \caption{a) Proton and b) electron energy spectra from the rear side of the target during an automated target position scan ($z_T$) with a \SI{12}{\micro\meter} Kapton tape and an on-target laser energy of $(438 \pm  32)\,\si{mJ}$.
    c) Average proton spectra (and standard deviation) for different $z_T$ positions as indicated in the legend.
    The proton spectra are recorded by the time-of-flight diamond detector.
    Each column of the waterfall plots is the average of the ten shots from each burst.
    The scan is comprised of 31 bursts at different target positions spaced at \SI{7.3}{\micro\meter} intervals along the laser propagation axis.
    Negative values of $z_T$ are when the target plane is closer to the incoming laser pulse and $z_T=0$ is the target at the best focus of the laser pulse.  The red data points, connected with a guide line, indicate the burst-averaged 95th percentile energy as well as the standard deviation on this value across the burst. 
    }
    \label{fig_waterfall}
\end{figure}

With the automated setup, parameter scans can be readily obtained by following a pre-programmed procedure.
In doing this, the control algorithm moved through a series of equally spaced locations, taking a number of repeat shots at each configuration to quantify shot-to-shot fluctuations.
Figure \ref{fig_waterfall} shows proton and electron spectra for a 1D scan of target position through the laser focus with a \SI{12}{\micro\meter} Kapton tape and a pulse length of $\tau_{\mathrm{FWHM}} =49\pm3\,\si{fs}$.
The burst averaged 95th percentile proton energy (hereafter referred to as the maximum energy) and average electron energies are overlaid in red with the standard deviation for each burst indicated by the error bars.
The proton and electron spectra are seen to extend to higher energies as the target position approaches the laser best focus as would be expected due to the increasing laser intensity at the target surface.
While the electron spectra peak around the best focus, where the laser intensity is highest, a characteristic dip in the maximum proton energy and flux is observed. 
Around best focus ($\left| z_T\right|<25\,\si{\micro\meter}$), a comparatively small number of protons are still observed at high ($\approx3.5\,\si{MeV}$) energy, with the spectrum dominated by lower (sub-MeV) energies, as seen in \cref{fig_waterfall}c.
A second signal is also seen in the TOF spectrum, appearing as a band peaking at $\approx 0.5\,\si{MeV}$ per nucleon in \cref{fig_waterfall}a.  This is most likely due to heavy ions that were accelerated to lower velocities due to their lower charge to mass ratio.
Together with the spectrally peaked proton spectrum, this appears similar to observations of `buffered' proton acceleration for higher intensities and thinner targets \cite{Dover2016NJP}, suggesting that the sheath field is dynamic during the acceleration process.

An increase in the laser focal spot size has previously been observed to increase the number of accelerated particles, although with a reduced maximum energy \cite{Green2010EnhancedIrradiation}. While this is consistent with our measurements, the strong suppression of proton flux at the highest intensity may indicate that the acceleration process is further compromised at the highest laser intensities by the contrast levels of our laser, with the prepulses and amplified spontaneous emission (ASE) causing adverse pre-heating of the target.
Similar disruption has previously been attributed to rear surface deformation by ASE-driven shock break-out, which can effectively steer a high energy component of the proton beam emission towards the laser axis \cite{Lindau2005,Lundh2007b}, modifying the spectrum measured at a single angular position, or the presence of a long scale length plasma on the rear-surface which has been shown to suppress the production of ions through TNSA in experiments and simulations \cite{Mackinnon2001EffectPulses,Fuchs2007Laser-foilGradients, Higginson2021InfluenceAcceleration}.

A laser pulse contrast measurement (using an Amplitude Sequoia) showed an ASE intensity contrast of better than $10^{-9}$ up to $t = -20\,\si{ps}$, after which the laser-intensity in the coherent pedestal \cite{Hooker2011OE} increased exponentially.  
Individual prepulses with a relative intensity of $10^{-6}$ were also observed between $t = -50\,\si{ps}$ and  $t = -65\,\si{ps}$. 
The measured contrast, starting at $t = -150\,\si{ps}$, was used to perform 2D cylindrical hydrodynamic modelling of the target evolution ahead of the arrival of the peak intensity.  The modelling was performed using the FLASH code (v4.6.2).  The ASE and coherent pedestal from $t = \SI{-150}{ps}$ to $t = \SI{-1}{ps}$ was coupled to the target electrons using ray-tracing with inverse bremsstrahlung heating, and Lee-More conductivity and heat exchange models were used.  This indicated the formation of an approximately \SI{2}{\micro\meter} exponential scale-length pre-plasma. For the target thicknesses used in the experiment ($\gg$~\SI{1}{\micro\meter}), the measured prepulse was not large enough for the generation of a shock moving quickly enough to perturb the density step of the target rear surface.
This matches previous results \cite{Kaluza2004InfluenceExperiments,Lundh2007b} at similar interaction conditions, which indicate that the ablation launched density shock does not have time to affect the rear surface during the acceleration process.
This dip in signal at highest intensities is a surprising result for targets with tens of microns thickness. It is more commonly observed for experiments using ultra-thin targets where it is attributed to  ASE shock-breakout \cite{Morrison2018MeVInteraction}.
For the interaction presented here, the rear surface may be affected by poor long-timescale contrast (before the start of the measurement window at $t = -150\,\si{ps}$), or through x-ray heating of the target bulk \cite{Kaluza2004InfluenceExperiments}.
Determination of the specific processes driving the disruption of proton acceleration for this interaction requires additional measurements of long-timescale contrast and pre-plasma scale length, and is beyond the scope of this optimisation demonstration.

 \begin{figure*}[htpb!]
    \includegraphics[width=17.4cm]{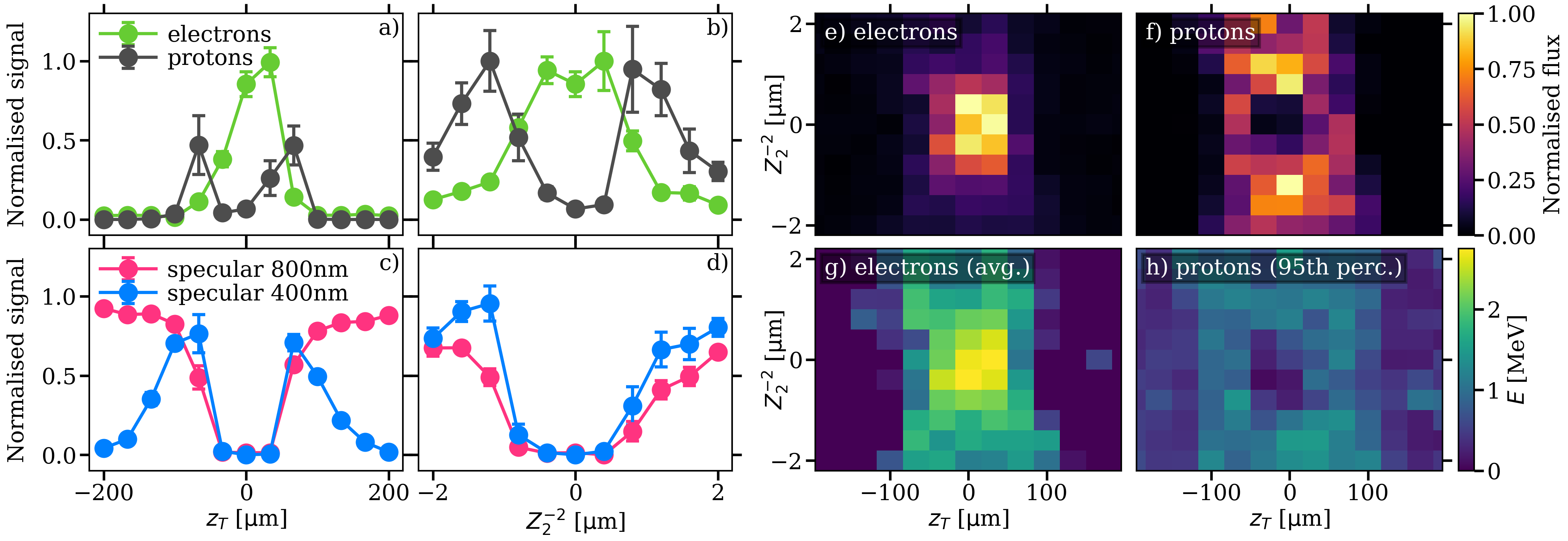}
    \caption{One dimensional scans of a) \& c) target z-position $z_T$ and b) \& d) astigmatism $Z_2^{-2}$ for \SI{12}{\micro\meter} thickness Kapton tape and a pre-plasma laser energy of $(453 \pm  40)\,\si{mJ}$.
    The electron and proton flux are plotted in a) \& b), and the specularly reflected fundamental and second harmonics laser signals are plotted in c) \& d).
    All fluxes are normalised to their observed maxima over the 2D parameter scans.
    Two dimensional scans of electron and proton flux are shown in e) \& f), with the average detected electron energy and the maximum (95th percentile) proton energies shown in g) and h) respectively.
    The 2D scan is a result of 143 bursts of 15 shots and the datapoints are the mean of each individual burst.
    %, with the standard error on the mean used as the error bars.
    }
    \label{fig:2dscans}
 \end{figure*}

The laser intensity can be varied by adding wavefront abberations to the laser, which changes the focal spot shape as well as the peak intensity. 
\Cref{fig:2dscans}a-d show the burst-averaged electron and proton flux as well as the relative specular reflectivity at the fundamental and second harmonic wavelengths for varying target plane, $z_T$ (\cref{fig:2dscans}a \& \ref{fig:2dscans}c), and 45-degree astigmatism, $Z_2^{-2}$ (\cref{fig:2dscans} b \& \ref{fig:2dscans}d).
Again, a characteristic drop was observed in the proton flux and maximum energy for best focus.
Moderately decreasing the peak intensity by either defocusing or adding astigmatism decreased the electron flux and average energy, but maximised the proton acceleration.
At the low intensity limit, the particle acceleration drops to zero as expected.

Evidence of disruption to the front surface during the interaction can be seen from the sharp drop in the fundamental and second harmonic laser reflectivity at high intensities.
This matches previous observations of target reflectivity and harmonic generation being adversely affected for high-intensity low-contrast laser-plasma interactions as a result of the formation of a large scale length pre-plasma \cite{Pirozhkov2009DiagnosticReflectivity,Streeter2011RelativisticGenerator,Chopineau2019IdentificationPlasmas}.

To explore the interplay between astigmatism and defocus, an automated 2D grid-scan was performed.
Burst averaged measurements of the electron and proton flux, mean electron energy and maximum proton energy are displayed in \cref{fig:2dscans} e-h. 
The electron generation was maximised for both zero defocus and zero astigmatism, monotonically decreasing as $z_T$ and $Z_2^{-2}$ were increased.
The proton flux and energy are approximately maximised for a ring around the origin, indicating a threshold intensity for disrupting the proton acceleration which can be achieved either by defocus or increasing the focal spot size through optical aberrations.
There also appears to be enhanced proton flux for zero defocus, $z_T=0\,\si{\micro\meter}$, but with the application of significant astigmatism  $Z_2^{-2}=\pm1.2\,\si{\micro\meter}$ when compared with the increased flux achieved just through defocusing with no astigmatism.
This indicates that the proton acceleration process is not just intensity dependent, but is also sensitive to the spatial intensity profile on-target \cite{Wilson2022InfluenceInteractions}.

\begin{figure}[h!]
    \includegraphics[width=0.5\textwidth]{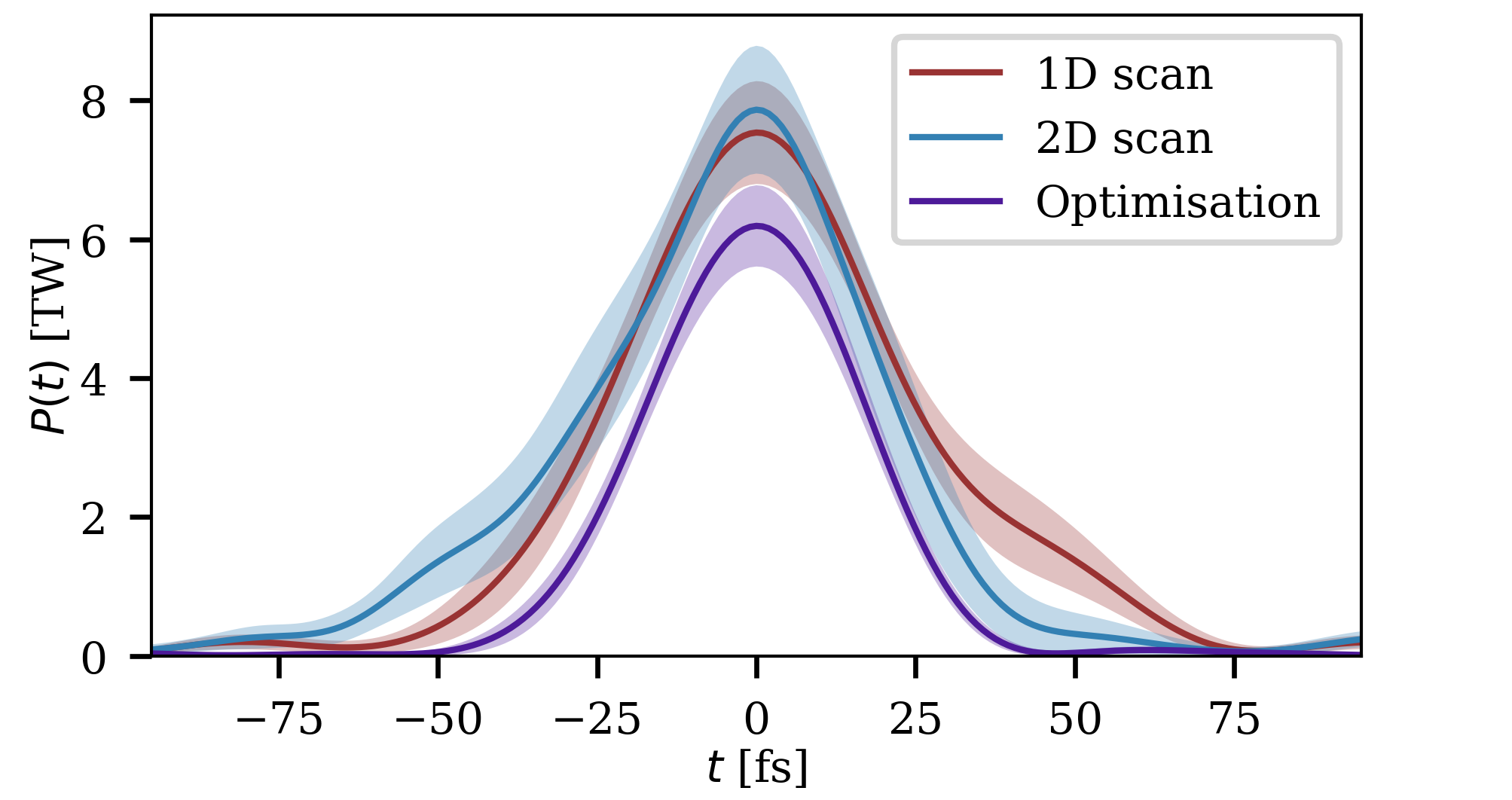}
    \caption{Laser pulses temporal profiles as measured by the on-shot SPIDER diagnostic for the results of the 1D scan (\cref{fig_waterfall}), 2D scan (\cref{fig:2dscans}) and optimisation (\cref{fig_opt_WF_PP}).
    The integrals of the signals are set by independent measurements of the on-target laser energy which were $(438 \pm  32)\,\si{mJ}$ (1D scan), $(453 \pm  40)\,\si{mJ}$ (2D scan) and $(258 \pm  22)\,\si{mJ}$ (optimisation).
    The corresponding measured FWHM pulse widths were $(49 \pm  3)\,\si{fs}$, $(45 \pm  4)\,\si{fs}$ and $(39 \pm  1)\,\si{fs}$. 
    }
    \label{fig_spider_pulses}
\end{figure}

For all results in \cref{fig:2dscans}, the laser pulse had a shorter pulse duration of $\tau_{\mathrm{FWHM}} =46\pm4\,\si{fs}$ and was skewed with a slower rising edge than for the results in \cref{fig_waterfall}, as can be seen in \cref{fig_spider_pulses}.
The pulse shape appears to effect the range of $z_T$ over which the proton acceleration was suppressed.
For the case of \cref{fig_waterfall}, proton acceleration was maximised at $z_T=\pm30\,\si{\micro\meter}$, while with the slower rising edge used for \cref{fig:2dscans}a, the maximum occurs at $z_T=\pm67\,\si{\micro\meter}$, with very low flux obtained at $z_T=\pm33\,\si{\micro\meter}$.
For a slower rising edge, there is more time for any disruption of the target to occur prior to the arrival of the peak of the pulse, meaning that a larger defocus is required to maintain proton acceleration.
A lower maximum proton energy of $0.8\pm0.1\,\si{MeV}$ was observed in \cref{fig:2dscans}a compared to $2.8\pm0.4\,\si{MeV}$ for \cref{fig_waterfall}.
This shows the benefit of using temporal pulse shaping to minimise target pre-heating, as it allows for efficient proton acceleration at higher intensity interaction leading to higher energy protons. Despite the very different interaction conditions (here - moderate laser contrast and micron thick target), this indicates a similar trend to that observed in recent experiments using high-contrast ultra-thin targets which demonstrated the enhancement of particle energies and numbers by modification of the pulse shape, away from nominal pulse compression, with a steepened rising edge in comparison to the falling edge of the pulse \cite{Ziegler2021ProtonSystem, Tayyab2018EffectTargets}.

The interaction parameter space is multi-dimensional. While 1D and 2D slices of this parameter space provide valuable insight, true mapping of the parameter space required for deeper understanding and control of sheath-accelerated proton beams, and the location of global optima, requires the inclusion of more dimensions.  
This is particularly important when individual parameters are coupled in complex relationships as is evident in the case of proton acceleration from \cref{fig:2dscans}.
While grid scanning is feasible for mapping up to 2D slices of parameter space with multi-Hz laser systems, for higher numbers of dimensions it becomes prohibitively time consuming.  Additionally, a grid-scan covers regions of parameters with high signal and no signal with the same resolution.  Given that measurements in large regions of the accessible parameter space will return no appreciable proton acceleration, this is uneconomical in terms of target usage, debris production and laser operation.
It is therefore desirable to use more intelligent algorithms for probing and optimisation of the beam in highly-dimensional parameter spaces.

\section{Bayesian Optimisation} \label{sec:results_BO}

In Bayesian Optimisation (BO) experimental data is used to update a \emph{prior} model to more accurately fit observations, thereby obtaining a \emph{posterior} model.
The model is then used to make predictions over the experimental parameter space and select the parameter set for the next measurement, with the goal of efficiently finding the optimum within the parameter space.
A commonly used type of model is Gaussian process regression (GPR) \cite{GaussianProcesses2005}, which is a well suited technique for modelling multi-dimensional experimental data. 
A key advantage for experimental science is that GPR can naturally include uncertainty quantification on the data, and also can be used to estimate the uncertainty when making predictions. 
BO is widely used for the optimisation of noisy processes that are expensive to evaluate and for which there is no adequate analytical description, as is typically the case in complex non-linear systems.

In the field of laser-plasma acceleration BO has been used in laser-wakefield acceleration to optimise the generated electron and x-ray beam properties \cite{Shalloo2020,Jalas2021} and in simulated laser-driven ion beams for maximising proton energy \cite{Dolier2022Multi-parameterSimulations}.
To demonstrate its applicability in a laser-driven ion acceleration experiment we have adapted the algorithm from Shalloo \emph{et al.} \cite{Shalloo2020}. 
Using the automated control of the experiment and on-line analysis of the experimental diagnostics we were able to perform multi-dimensional optimisation of any fitness function which outputted a scalar property of interest, such as the maximum proton energy.

The input values passed to the model for each burst were taken from the parameters set by the control algorithm with the exception of the plane of the interaction, $z_T$.
It was found that errors in positioning the tape target, while significantly smaller than the Rayleigh range of the focusing laser (\SI{15}{\micro\meter}), were large compared to the sensitivity of the plasma accelerator and so the target position input values were taken from a spatially resolved measurement of the self-emission region at the target surface, collected at $60^{\circ}$ to the front surface normal of the tape.
This was found to greatly improve the confidence of the model and its ability to find the optimum.

\begin{figure}[h!]
    \includegraphics[width=8.4cm]{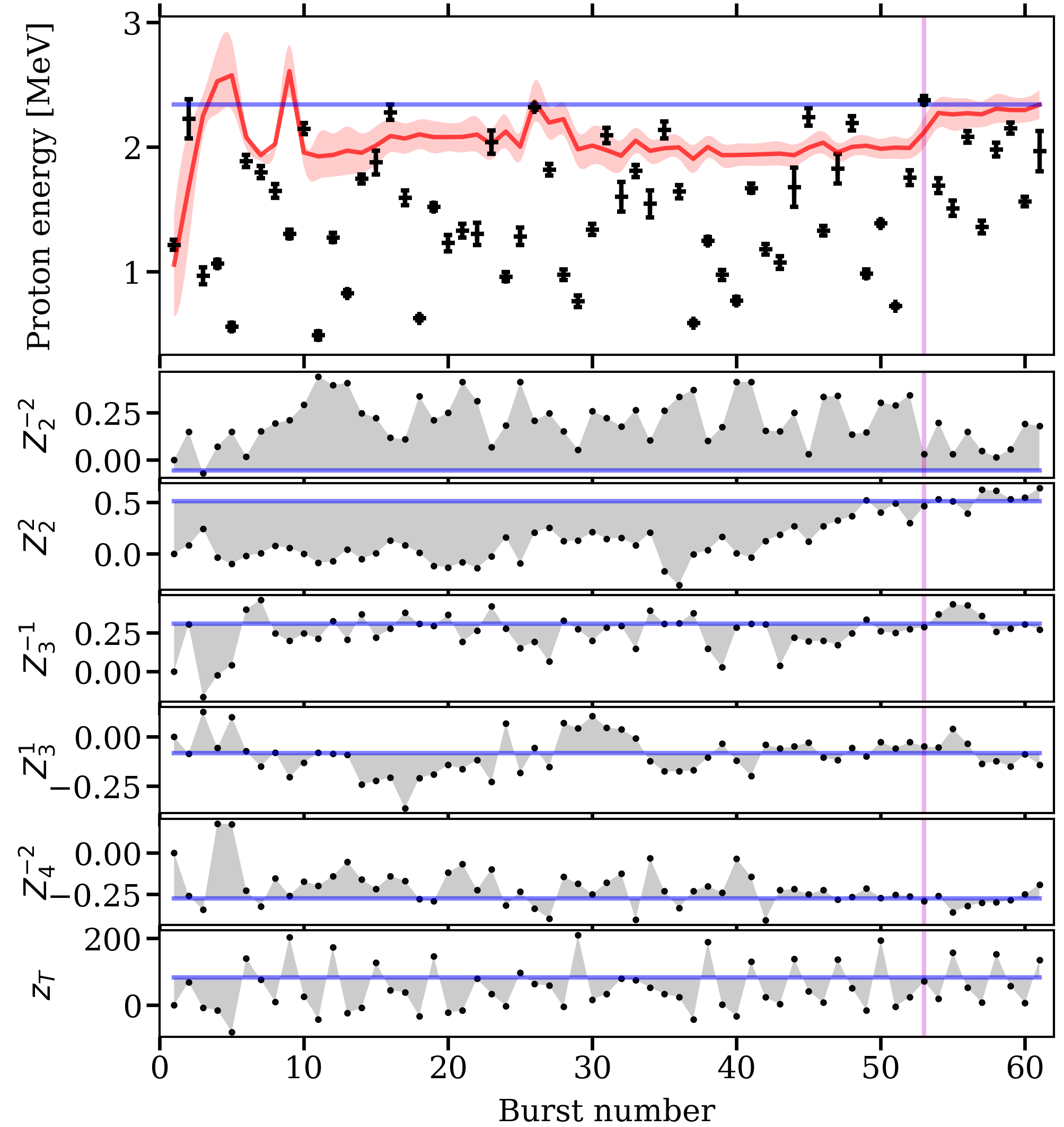}
    \caption{Optimisation of the 95th percentile proton energy determined by the rear surface time-of-flight diagnostic through the adjustment of the laser wavefront and position of target along the laser propagation direction ($z_T$).
    The top panel shows the measured values of the proton energy (median and median absolute difference of each burst) as a function of the burst number (black points and error bars respectively), together with the model predicted optimum after each burst (red line and shaded region) as well as the final optimal value from the model (blue horizontal line).
    The variation of each control parameter (given in microns) is shown in the lower plots (black points) along with the final optimised values (blue horizontal line), also as functions of burst number.
    The best individual burst is indicated by the vertical magenta line in each plot and it can be seen that for all parameters the experimental parameters fall very close to the optimum value predicted by the model (e.g. they are close to the horizontal blue line). 
    %Each burst contains 20 shots and the proton energy data points are the mean and standard error of each individual burst.
    For this data series, each burst contained 20 shots, the target was \SI{12}{\micro\meter} Kapton tape and the laser energy was $258 \pm 22\,\si{mJ}$.}
    \label{fig_opt_WF_PP}
\end{figure}

To demonstrate the BO algorithm, a 6D optimisation was performed to maximise the maximum proton energy (as measured by the TOF).
Five Zernike mode coefficients - including $Z_2^{-2}$, $Z_2^{2}$ (oblique \& vertical astigmatism), $Z_3^{-1}$, $Z_3^{1}$ (vertical \& horizontal coma) and $Z_4^{-2}$ (oblique second astigmatism) -
%$Z_1^{-1}$, $Z_1^{1}$, $Z_2^{-2}$, $Z_2^{2}$, $Z_4^{0}$ 
were used to change the spatial phase of the laser pulse, affecting the focal spot shape and peak intensity.
In addition, the tape surface position relative to the focal plane of the laser $z_T$, was varied using a linear motorised stage and an on-line self-emission diagnostic to measure its position, %with a precision of \SI[2]{$\mu$m} 
as mentioned.
This optimisation started from an initially flat wavefront (optimised using a feedback loop with a HASO wavefront sensor) and the tape target was initially positioned at the estimated focal plane of the laser.  This represents the typical starting point for conventional optimisation of manual experiments, with the highest intensity being assumed to be optimal.

\Cref{fig_opt_WF_PP} shows the measured maximum proton energy along with the variation of each input parameter as a function of burst number.
By chance, one of the initial randomly selected points produced a large enhancement in maximum proton energy with every parameter apart from $Z_2^2$ close to its eventual optimum value.
With each additional measurement, the model gained more knowledge of the parameter space and adjusted its prediction of the global optimum of proton energy (red line) and its location in parameter space.
After 61 bursts, the model optimum was $(2.30 \pm 0.10)\,\si{MeV}$, compared to a starting point of $(1.22 \pm 0.04)\,\si{MeV}$.
This optimised maximum energy is close to the value shown in \cref{fig_waterfall} and significantly greater than seen in \cref{fig:2dscans}, despite being limited to only \SI{260}{mJ} of on-target laser energy, compared to $>430\,\si{mJ}$ for the parameter scans.

\begin{figure}[h!]
    \includegraphics[width=8.4cm]{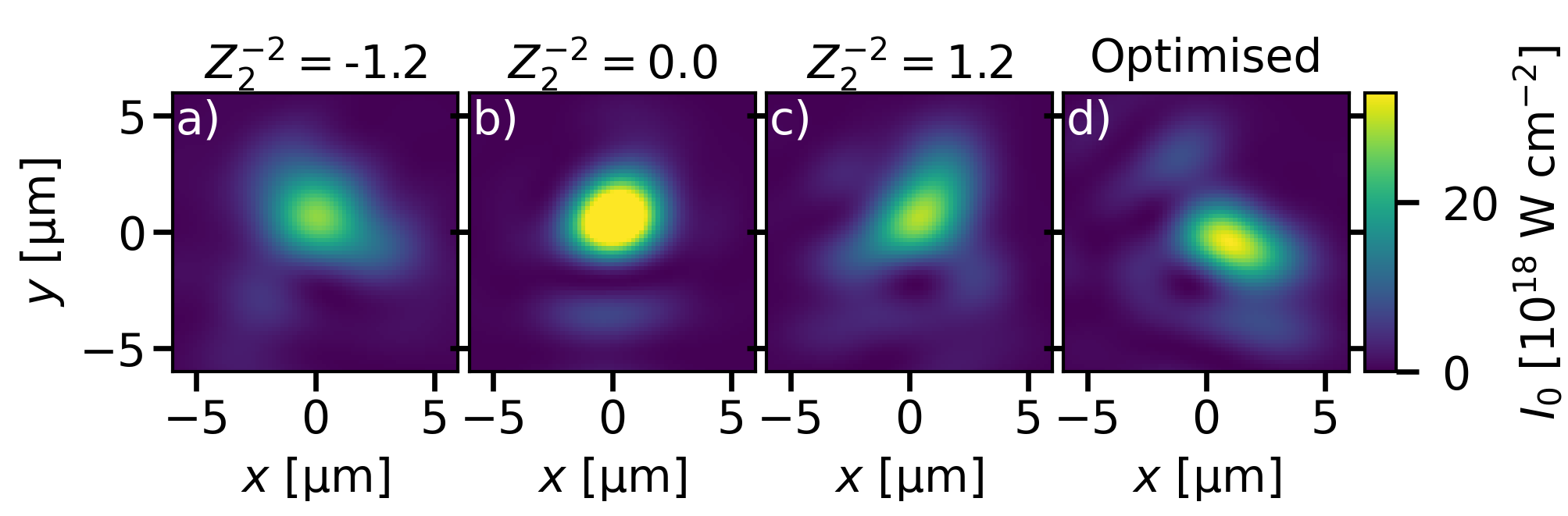}
    \caption{Reconstructed laser intensity profiles at $z_T = 0\,\si{\micro\meter}$ for  a) $Z_2^{-2}=-1.2\,\si{\micro\meter}$, b) $Z_2^{-2}=0.0\,\si{\micro\meter}$ c) $Z_2^{-2}=1.2\,\si{\micro\meter}$ and d) for the optimal pulse (burst 53) from the optimisation shown in \cref{fig_opt_WF_PP}.
    The peak intensity of each focus was (2.7, 5.1, 2.9 and 3.2) $\times 10^{19}$ \,{Wcm$^{-2}$} respectively.}
    \label{fig_optimised_focus}
\end{figure}

The optimum found involved a shift of \SI{70}{\micro\meter} from the initial position (best focus), as well as the addition of significant wavefront aberrations compared to the initially flat wavefront.
The focal spot fluence distribution at the target plane was calculated from on-shot measurement of the near-field phase and fluence profiles using a HASO wavefront sensor.
The resulting intensity maps are shown in \cref{fig_optimised_focus} for $Z_2^{-2} = -1.2,0,1.2$~\SI{}{\micro\meter} at zero defocus and for the optimised wavefront found through the optimisation.
The peak intensity was $I_0 = 5\times 10^{19}\,\si{W cm^{-2}}$ for the case of a flat wavefront at the focal plane.
Each of the modified pulses had a similar peak intensity of $I_0 \approx 3\times 10^{19}\,\si{W cm^{-2}}$, with the spot shape appearing close to an ellipse for the optimised pulse case.
From analysis of the previous parameter scans it is inferred that this intensity distribution was optimal maintaining a maximum possible intensity, while limiting disruption to the rear target surface.

The reason for the focal spot shape found in this optimisation would require expensive 2-3D numerical simulations investigating how subtle changes in wavefront affect the various energy transfer processes and plasma dynamics in sheath acceleration.
This would be valuable for understanding how to further optimise laser-driven proton acceleration, but is beyond the scope of this paper which is focused on the utility of online optimisation for proton beam parameter optimisation as a particle source for applications as well as its use to identify interesting regions for deeper study within a complex non-linear system.

\section{Conclusion} \label{sec:conclusion}

In conclusion, we have demonstrated the automation and optimisation of laser-driven proton acceleration from a solid tape drive.
The ability to take high fidelity parameter scans in one or two dimensions will be of great benefit to the field in understanding what is a highly complex and dynamic interaction.
Bayesian optimisation of the generated proton beam was demonstrated, by using real-time analysis of experimental diagnostics to create a closed-loop system with limited human intervention.
This can find optima that would elude manual optimisation or single parameter scans, due to the complex interplay between the large number of control parameters.
In the case of this low-contrast interaction, the temporal pulse shape was shown to play an important role in determining the intensity threshold at which the proton acceleration process was disrupted.
Including temporal and spatial pulse shaping simultaneously in the optimisation process may lead to further improvement.

Automated Bayesian optimisation can quickly find regimes of stable and optimised operation without requiring the constant attention of laser-plasma experts.
It is anticipated that this development will be essential for efficient utilisation of laser-driven ion acceleration for its many applications in future user facilities \cite{qubs6040030}.
Bayesian optimisation could also be extremely valuable for optimising  radiation pressure acceleration for which laser-plasma instabilities \cite{Palmer2012PRL,Sgattoni2015PRE} typically limit the acceleration process, as well as for optimising enhanced acceleration through relativistic transparency, which has already demonstrated highly desirable near-~\SI{100}{MeV} proton energies \cite{Higginson2018Near-100Scheme} and for which target evolution plays a central role.
Fine tuning of the laser parameters may be able to mitigate these instabilities or further tailor the target evolution respectively, significantly enhancing the accelerated proton beam.

\vspace{10pt}
\section*{Data Availability Statement}

The data that support the findings of this study are available from the corresponding author upon reasonable request.

\section*{Acknowledgements} \label{sec:acknowledgements}
 
    We acknowledge support from the UK STFC grants ST/V001639/1 with the XFEL Physical Sciences Hub and ST/P002021/1,  the UK EPSRC grants EP/V049577/1 and EP/R006202/1, as well as the U.S. DOE Office of Science, Fusion Energy Sciences under FWP No.~100182, and in part by the National Science Foundation under Grant No.~1632708 and Award No.~PHY~–~1903414.
    M.J.V.S. acknowledges support from the Royal Society URF-R1221874.
    G.D.G. acknowledges support from the DOE NNSA SSGF program under DE-NA0003960. 
    A.G.R.T acknowledges support from the US DOE grant DE-SC0016804.
    D.M. acknowledges support from the project `Advanced research using high-intensity laser-produced photons and particles (CZ.02.1.01/0.0/0.0/16\_019/0000789)' from European Regional Development Fund (ADONIS)

   Special thanks goes to the staff at the Central Laser Facility who provided laser operational support, mechanical and electrical support, computational and administrative support throughout the experiment.

\bibliography{references}

\end{document}